# Continuum Damage Modeling of Biaxial Fatigue Failure in Whole Bone: A Hybrid Approach with Machine Learning Integration

Reza Kakavand* [1,2,3], Jonah M. Dimnik[1,2,3], Ifaz T. Haider[1,2,3], W. Brent Edwards[1,2,3]

[1] Human Performance Laboratory, Faculty of Kinesiology, University of Calgary, Calgary, Canada

[2] McCaig Institute for Bone and Joint Health, University of Calgary, Calgary, Canada

[3] Department of Biomedical Engineering, Schulich School of Engineering, University of Calgary

**\* Corresponding author:**

**Reza Kakavand, PhD, (*He/Him*)**

reza.kakavand@ucalgary.ca

**Postdoctoral Fellow,**

**HPL, Faculty of Kinesiology**

**University of Calgary**

**2500 University Drive NW**

**Calgary, AB, Canada T2N 1N4**

**Abstract**

Repetitive loading of bone is associated with microdamage accumulation and material property degradation that may ultimately result in fatigue fracture. Our previous work used continuum damage mechanics (CDM)-based finite element (FE) modeling to predict stiffness loss and fatigue failure in whole bone; however, this model did not account for inter-specimen variability in fatigue behaviour and multiaxial loading effects, which limited its applied efficacy. In this study, we refined the CDM-based FE model to predict experimental fatigue-life measurements from 21 whole rabbit tibiae subjected to cyclic axial compression with different magnitudes of superposed torsion. Machine learning (ML) methods were used to predict damage parameters from initial, undamaged, conditions, which were then used to predicted stiffness degradation and fatigue failure with the CDM-based FE modeling pipeline. Regression analysis was used to compare CDM-based FE fatigue-life predictions with experimental measurements. A random forest ML model predicted specimen-specific damage parameters with high accuracy ($R^2 = 0.85$) and the CDM-based FE models demonstrated remarkable predictive capability, explaining up to 91% of the variance in fatigue-life measurements. Stiffness degradation profiles also followed a similar trend to experimental measurements; however, this agreement worsened with the level of superposed torsion suggesting additional refinements to the model may be necessary. These findings demonstrate the efficacy of integrating ML with CDM-based FE modeling to predict fatigue life and stiffness degradation. The observed agreement with experimental measurements suggests the modeling framework may provide valuable information regarding the mechanisms of fatigue fracture in whole bone.

# 1  Introduction

Repetitive loading of bone is associated with microdamage accumulation, corresponding with localized material property degradation and a progressive loss in structural stiffness [1–4]. This process, known as mechanical fatigue, may eventually lead to fracture at loading magnitudes much lower than the monotonic strength of bone [5,6]. Mechanical fatigue is believed to play an important role in stress fractures in athletes and military personnel [7–10], as well as insufficiency fractures in older adults and those with metabolic bone disease [11–14]. Thus, an understanding of the mechanical fatigue behavior of bone has clinical relevance and may be crucial for developing strategies to prevent fatigue-related fractures and optimize rehabilitation approaches [15–18].

The mechanical fatigue behavior of bone has traditionally been studied at the material-level, by cyclically testing small, machined samples *ex vivo* [19–23]. These tests have provided valuable information about the relationship between the fatigue life (i.e., number of cycles to failure) of bone tissue and attributes such as loading magnitude [19,24], loading mode [25,26], and bone microarchitecture [1,21–23], among others. Fatigue tests of whole-bone are conducted less frequently [27,28], in part, because meaningful conclusions about loading conditions and bone properties are complicated by localized stress/strain states arising from specimen-specific geometry and heterogeneous material (i.e., density) distributions [29,30]. In this regard, finite element (FE) modeling has emerged as a powerful tool for estimating the micro-mechanical environment of whole bones under load [27,31].

Previous work, employing a "stress-life approach", illustrated strong relationships between FE predictions of the initial strain distribution and the fatigue life of whole bone [27,31,32]; however, mechanical fatigue is a dynamic, nonlinear process, and few studies have explicitly considered the degradation of material properties associated with fatigue damage, and how this may influence fatigue-life measurements. [32,33] We previously introduced a continuum damage mechanics (CDM)-based FE model to simulate fatigue damage and failure in whole rabbit tibiae [33]. The rabbit is a commonly used model in orthopaedics because of its manageable size, natural Haversian remodeling, and well-established surgical techniques [34–36]. The

CDM-based FE model explained 71% of the variability in experimentally measured fatigue life; however, the model was calibrated using only uniaxial loading, and a single set of optimized parameters was applied across all samples, which did not account for inter-specimen variability in fatigue resistance [19,37,38]

In this study, we introduce a CDM-based FE model to simulate cyclic biaxial loading with inter-specimen considerations for damage accumulation. The study had three main objectives. First, the CDM-based FE modeling pipeline was used to optimize the damage parameters for each whole rabbit tibia to achieve less than 1% error in experimental fatigue-life measurements (referred to as the true damage parameters). Secondly, we leveraged Machine Learning (ML) to study the prediction of damage parameters from initial, i.e., undamaged, mechanical behaviour, and their interactions. Lastly, the ML-predicted damage parameters were incorporated into the CDM-based FE modeling pipeline to predict fatigue life. The observed agreement with experimental measurements suggests the modeling framework can provide valuable information regarding the mechanisms of fatigue fracture in whole bone.

## 2 Material and methods

The following sections outline the experimental setup used for model valuation, CDM-based FE modeling pipeline for damage parameters optimization, damage model formulation, and ML implementation used in this study.

### 2.1 Optimization of Damage Parameters for Rabbit Tibia under Cyclic Biaxial Loading

#### 2.1.1 Experimental setup

The experimental data used in this study were obtained from a previous investigation of whole rabbit tibiae exposed to cyclic biaxial loading [27]. In that study, tibiae from female New Zealand White Rabbits were extracted post-mortem and scanned using a Discovery 610 CT scanner (General Electric Healthcare; WI, USA) with the following acquisition parameters: tube current 220 mA, peak voltage 120 kVp, pitch = 1, rotation time 0.5 s, in-plane voxel size of 0.39 mm, and between-plane voxel size of 0.625 mm. A calibration phantom with known hydroxyapatite equivalent bone mineral density inserts were included in the CT scan field-of-view (QRM GmbH; Möhrendorf, Germany). The proximal and distal most 1 cm of each bone were then potted in stainless steel fixtures for mechanical testing. Fatigue experiments were conducted under cyclic axial compression combined with varying levels of torsion. Specifically, bones were tested at 50% of their ultimate compressive force (1260 N) with one of three different torsional loading conditions: 0% (n = 10), 25% (n = 6), and 50% (n = 5) of ultimate torsion (3.4 Nm). For each specimen, loading was applied at 2 Hz until failure with the minimum absolute load held fixed at 20 N, such that the load ratio, min/max load magnitude, was always close to zero. This dataset provided a well-controlled experimental foundation for evaluating fatigue behavior under physiologically relevant loading conditions, enabling further computational modeling and ML analysis in the present study (Figure 1).

#### 2.1.2 Finite element (FE) modeling

Specimen-specific FE models were developed for each rabbit tibia based on the CT scans obtained prior to mechanical testing. The segmentation of bone from the background image was performed using a density threshold of 0.12 g/cm³ [27]. The segmented bone volume was then meshed using second-order tetrahedral elements with an average edge length of 0.8 mm, resulting in models containing approximately 70,000 elements per tibia. A mesh convergence study confirmed that increasing the number of elements threefold altered peak strain values by less than 2%, indicating sufficient resolution for FE analysis.

The material properties of the bone were assigned using a heterogeneous, anisotropic, linear elastic formulation. Young's modulus ($E_3$) in the axial direction was determined at each element location based on its bone mineral density using the relationship [29]:

$$E_3 = 8105 \cdot \rho_{HA}^{1.75} \tag{1}$$

where $\rho_{HA}$ represents the hydroxyapatite (HA) equivalent bone mineral density (g HA/cm³) established with the calibration phantom. This relationship was recently shown to provide excellent agreement (i.e., $R^2$ = 0.964) between experimentally measured and FE-predicted principal strains for the rabbit tibia [29]. Anisotropy was defined as $E_1=0.574 \times E_3$, $E_2=0.577 \times E_3$, $G_{12}=0.195 \times E_3$, $G_{23}=0.265 \times E_3$, $G_{31}=0.216\, E_3$, $\nu_{12}=0.427$, $\nu_{23}=0.234$, and $\nu_{31}=0.405$ in accordance with human cadaveric torsional tests [39,40]. A preliminary sensitivity analysis showed anisotropic FE models exhibited a significant reduction in fatigue life with increasing superposed torsion ($p < 0.05$), similar to the experimental data, while isotropic FE models exhibited no significant differences among groups (Figure S1 in the **supplementary material**).

Boundary conditions were applied by constraining the distal 1 cm of the bone in all degrees of freedom, while the proximal end was restricted in transverse translation but allowed to rotate around the long axis and translate axially. The applied loads—axial compression with or without torsion—were distributed across the nodes of the proximal region, matching the loading conditions used in the experimental tests. All FE simulations were conducted using ABAQUS software (version 2020, Dassault Systèmes, RI, USA).

2.1.3 Continuum Damage Mechanics (CDM)

A damage model (Eq. 2) was adapted from the damage rate formulation [19]:

$$\Delta D = \Delta N \times A \times \varepsilon_{vm}{}^B \tag{2}$$

where $\Delta N$ is the step time, A and B are damage parameters (constants), and $\varepsilon_{vm}$ is von Mises equivalent strain. The value of the time step parameter, $\Delta N$, was selected to be a reasonable compromise between solution accuracy and computational time. The time step parameter can take any value, and it represents the number of loading cycles between each damage calculation (i.e., for $\Delta N = 1000$, each FE analysis iteration would represent 1000 loading cycles, and the damage is calculated as 1000 times the damage incurred during a single loading cycle). In this study, the time step parameter was set to 10 for samples in the low-cycle regime (fatigue life < $10^3$), 100 for those in the medium-cycle regime ($10^3 \leq$ fatigue life < $10^4$) and 1000 for those in the high-cycle regime (fatigue life $\geq 10^4$). The von Mises equivalent strain was calculated according to the ABAQUS formulation in Eq. (3):

$$\varepsilon_{vm} = \sqrt{\frac{2}{3} \varepsilon'_{ij} \varepsilon'_{ij}} \tag{3}$$

$$\varepsilon'_{ij} = \varepsilon_{ij} - \frac{1}{3} \delta_{ij} \varepsilon_{kk} \tag{4}$$

where $\varepsilon_{ij}$ is the total strain tensor, $\delta_{ij}$ is Kronecker delta, and $\varepsilon_{kk}$ represents the volumetric strain.

We evaluated several additional strain-based criteria for the damage evolution law, including maximum shear strain, maximum principal strain, and strain energy density (SED). All showed similar variability in parameters A and B across samples, but we ultimately chose von Mises equivalent strain for consistency with our previous work [33].

### 2.1.4 Optimization of damage parameters

The damage constants A and B were determined through an optimization routine designed to minimize the fatigue-life error to be less than 1% (referred to as the true damage parameters). Two scenarios were considered: in the first, B was held constant at 3.241 [33], while A was calibrated; in the second, A was fixed at 10.40 [33], while B was calibrated. The von Mises stress distribution, max principal strain distribution, damage distribution and compressive stiffness degradation were compared between the two scenarios.

After calculating the damage evolution at a given number of cycles, we used the damage value to estimate the elastic modulus degradation at each integration point according to Eq. (5):

$$\tilde{E}_3 = E_3^0 (1 - D) \tag{5}$$

where, $\tilde{E}_3$ represents the new elastic modulus after incurring damage and $E_3^0$ is the initial, undamaged, elastic modulus; anisotropy, as described above, was assumed to be constant throughout the simulation. A 25% reduction in whole-bone compressive stiffness was assumed as the failure criterion in accordance with previous work [33]. The damage process was implemented in a user material subroutine (UMAT) in ABAQUS.

## 2.2 Machine Learning Prediction of Damage Parameters

Inter-specimen variability in damage parameters, as shown in previous studies [19,33], suggests the need for a specimen-specific damage model. In principle, specimen-specific damage parameters reflect differences in intrinsic and extrinsic toughening mechanisms over the course of the fatigue failure process. We leveraged ML to predict the damage parameters, A and B from Eq 2. Three ML models—Linear Regression, Random Forest (RF), and Support Vector Regression (SVR)—were implemented using their default parameters from the Scikit-Learn Python package. Given the limited dataset size (n = 21), leave-one-out cross-validation (LOOCV) was used as the model evaluation technique. In LOOCV, each sample is iteratively used as the test set while the remaining (n-1) samples serve as the training set. This approach ensures that all data points contribute to both model training and validation, making it particularly suitable for small datasets like the one in this study [41].

The input features used for training included bone mineral content (BMC), initial compressive stiffness, initial torsional stiffness, and their combinations. The initial mechanical parameters were obtained from the FE model, with the rationale that they could be easily obtained for new samples where the fatigue behavior was unknown. A comprehensive sensitivity analysis was conducted to evaluate the impact of different feature combinations on model performance, ensuring that the selected models were robust and capable of capturing key relationships between input parameters and the damage parameters A and B from Eq 2. Among these features, initial compressive stiffness and initial torsional stiffness showed the best predictions. Therefore, these features were selected for further analysis.

## 2.3 Specimen-Specific Fatigue Life Prediction

The ML-predicted damage parameters in section 2.2 were incorporated into the CDM-based FE modeling pipeline to predict fatigue life.

## 2.4 Statistical Analysis

Regression analysis was conducted to evaluate the relationship between true damage parameters and experimental fatigue life. Power-law relationships were explored, and the coefficient of determination ($R^2$) was reported. Statistical differences in predicted fatigue life across loading groups were assessed using the Kruskal-Wallis test, following a Shapiro-Wilk test for normality. ML performance was evaluated using the coefficient of determination ($R^2$), root mean squared error (RMSE) and mean absolute error (MAE). Additional regression analysis was performed to compare FE fatigue-life predictions with experimental measurements.

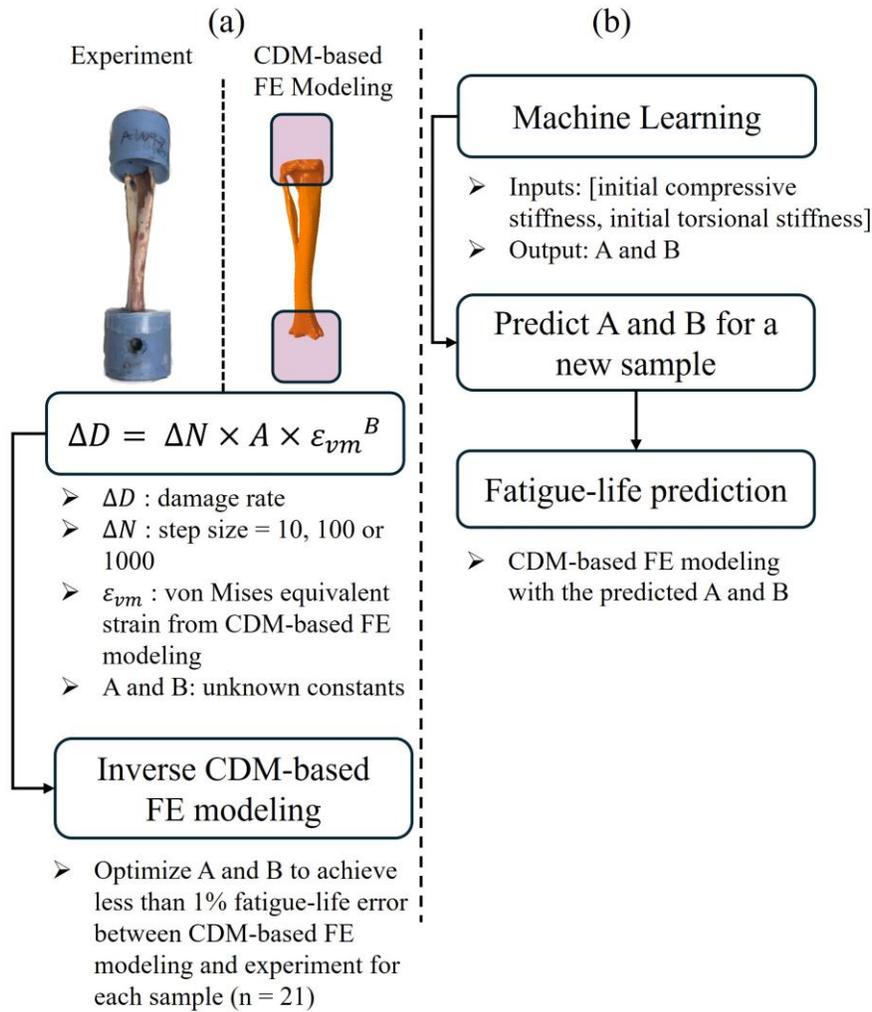

Figure 1. Workflow of the modeling pipeline: a) The true damage parameters (A and B) were computed for each sample, and b) Machine Learning models were trained using initial compressive stiffness, and initial torsional stiffness from FE analysis to predict damage parameters (A and B from Eq 2). The predicted damage parameters were then used in CDM-based FE modeling to predict fatigue life for each sample (n = 21).

# 3   Results

The optimized values for the true damage parameters A and B (Eq 2) are listed in Table 1. Parameter A illustrated a decreasing trend with the number of cycles to failure, while parameter B illustrated an increase (Figure 2a and 2b). No statistically significant differences in A or B were observed between loading groups, as confirmed by the Kruskal-Wallis test (Figure 2c and 2d). This indicated that variation in parameters A and B was associated with differences in bone mechanical properties rather than loading conditions. The 3D distributions of stress, strain, and damage revealed identical patterns for both calibration methods, i.e., either holding A or B constant (Figure 3). Agreement between experimental measurements and model predictions can also be observed by examining degradation in compressive stiffness prior to failure (Figure 4). It is interesting to note that model predictions became closer to the upper bound of experimental measurements as the level of superposed torsion increased.

Overall, the ML models predicted B with greater accuracy than A (Figure 5). Among the ML models, Support Vector Regression provided the best prediction for A ($R^2$ = 0.657), while Random Forest provided the most accurate prediction for B ($R^2$ = 0.854), based on $R^2$, RMSE, and MAE. The A and B values predicted by ML (i.e., Support Vector Regression for A and Random Forest for B) were used in CDM-based FE models to predict cycles to failure (Figure 6). The ML models trained on B explained 91% of fatigue-life variance, compared to 85% when trained on A.

Table 1. Calibrated damage parameters, A and B from Eq 2.

| Torsion | Sample number | A (B=3.241) | B (A=10.40) |
|---|---|---|---|
| 0% | 1 | 25.0 | 3.035 |
| | 2 | 23.0 | 3.040 |
| | 3 | 70.0 | 2.730 |
| | 4 | 3.5 | 3.540 |
| | 5 | 133.2 | 2.595 |
| | 6 | 40.0 | 2.925 |
| | 7 | 2.2 | 3.607 |
| | 8 | 6.5 | 3.330 |
| | 9 | 9.5 | 3.265 |
| | 10 | 38.0 | 2.907 |
| 25% | 11 | 3.7 | 3.487 |
| | 12 | 10.4 | 3.253 |
|     | 13   | 13.3  | 3.181 |
|     | 14   | 25.3  | 3.013 |
|     | 15   | 87.5  | 2.695 |
|     | 16   | 135.0 | 2.540 |
| 50% | 17   | 35.0  | 2.930 |
|     | 18   | 36.0  | 2.884 |
|     | 19   | 28.5  | 2.980 |
|     | 20   | 61.0  | 2.730 |
|     | 21   | 93.0  | 2.622 |
|     | mean | 40.1  | 3.017 |

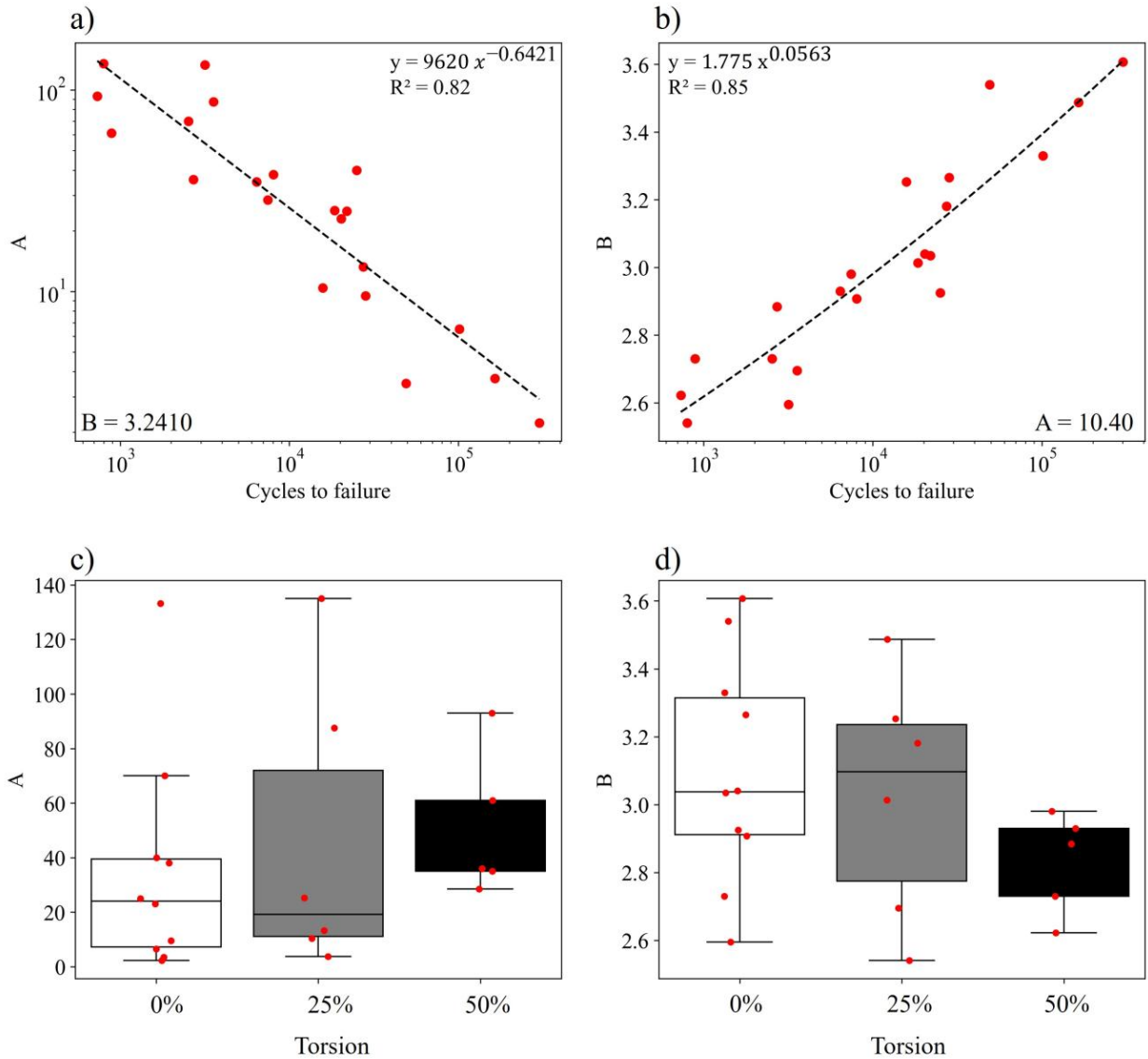

Figure 2. Calibrated damage parameters A and B from Eq 2. a) Calibrated A (B = 3.241) and b) calibrated B (A = 10.40) against cycles to failure. c) Calibrated A (B = 3.241) and d) calibrated B (A = 10.40) as a function of loading group, i.e., relative amount of superposed torsion.

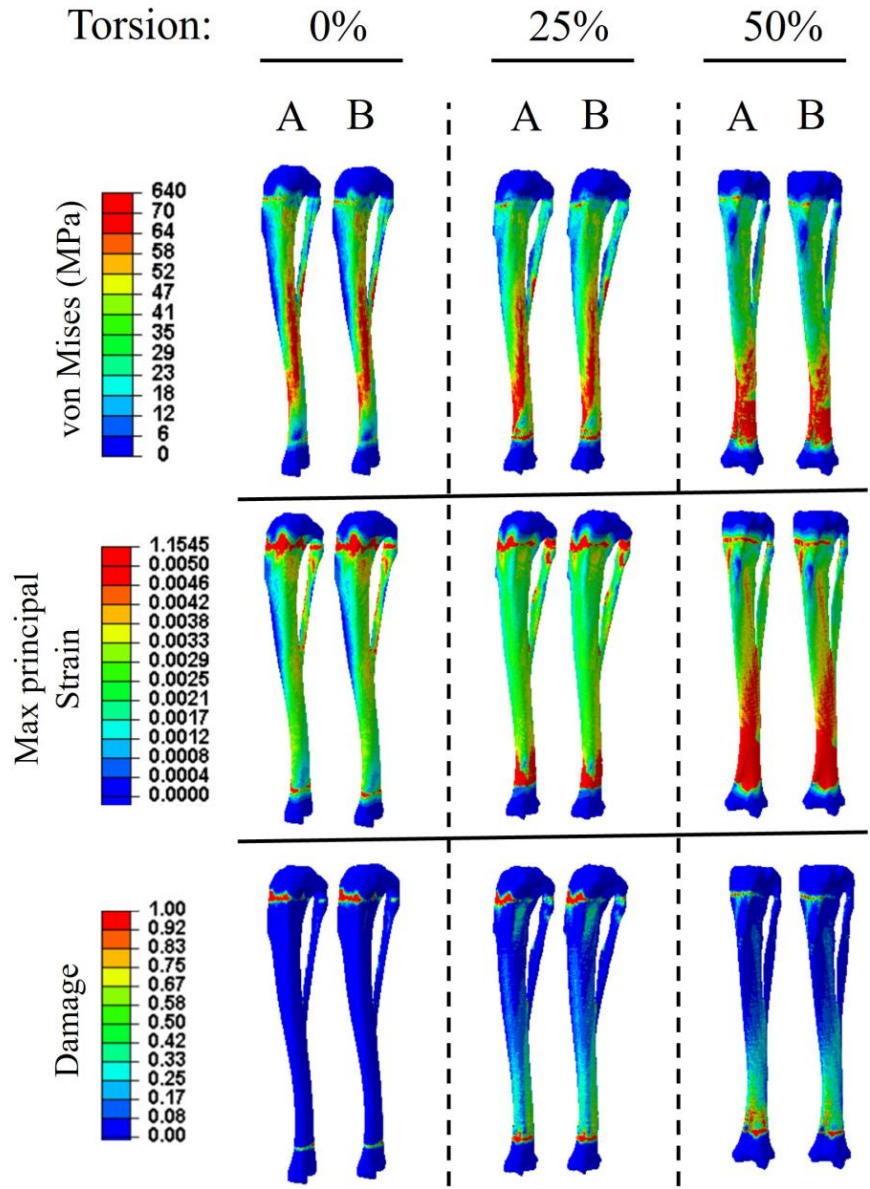

Figure 3. The von Mises stress (MPa), max principal strain and damage distributions for three representative samples in different loading groups. Columns with the header A indicate the models optimized for A (B = 3.241) and columns with B indicate the models optimized for B (A = 10.40).

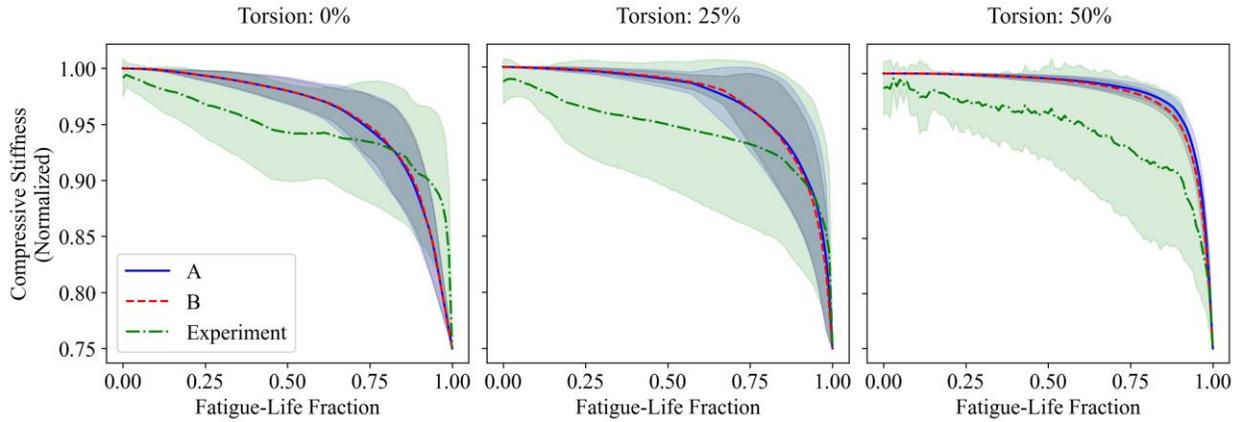

Figure 4. Normalized compressive stiffness against the fatigue-life fraction across all samples (n = 21). The solid lines show simulations where A was calibrated (B = 3.241), dashed lines show simulations where B was calibrated (A = 10.40) and dash-dotted lines shows the experimental measurements.

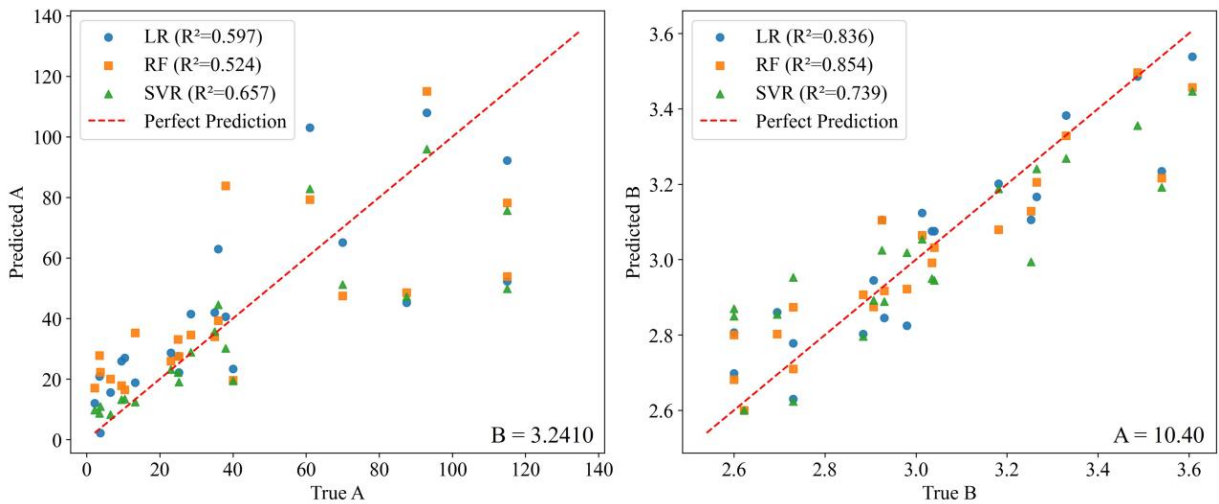

Figure 5. Predicted A (B = 3.241) and predicted B (A = 10.40) by ML models against true A and B (optimized in this study) from Table 1. LR: Linear Regression, RF: Random Forest, SVR: Support Vector Regression.

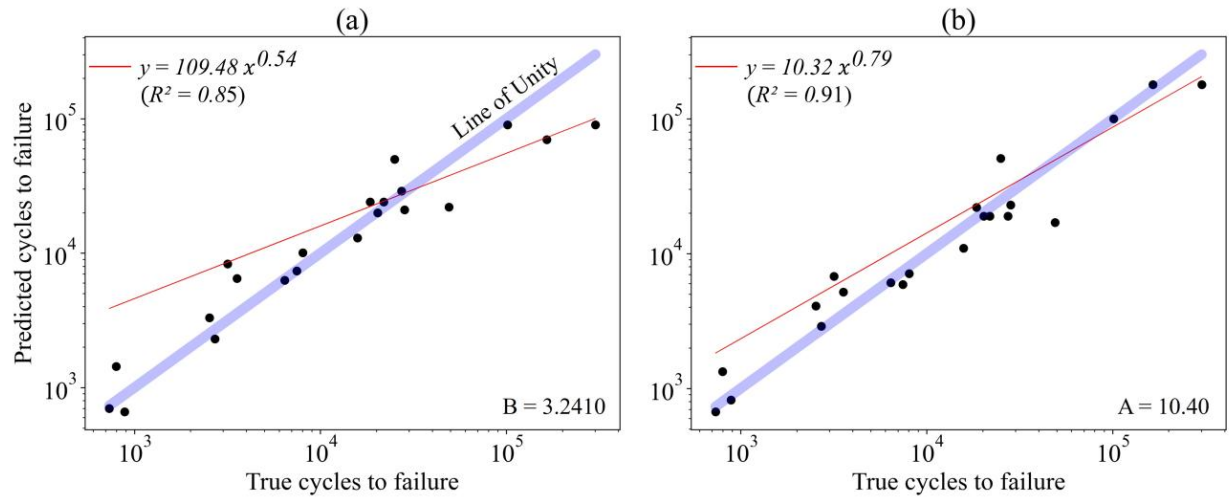

Figure 6. CDM-based FE predicted verses experimentally measured cycles to failure. a) Calibrated on A (B = 3.241), and b) calibrated on B (A = 10.40).

## 4 Discussion:

The purpose of this study was to develop and validate a CDM-based FE model to simulate cyclic biaxial loading with inter-specimen considerations for damage accumulation. Three objectives were addressed. The first objective utilized experimental fatigue-life data from 21 whole rabbit tibiae under cyclic biaxial loading in a CDM-based FE model to obtain damage parameters (A, B from Eq. 2) for each bone. We referred to these as true damage parameters since they were optimized to result in CDM-based FE-predicted fatigue life within 1% of the experimental fatigue life. The second objective employed LOOCV ML models to predict specimen-specific damage parameters for each sample. Finally, the third objective used the ML predicted damage parameters in the CDM-based FE model to predict fatigue-life measurements. The results demonstrated the strong potential of a CDM-based FE approach, informed by ML, for predicting stiffness degradation and ultimately fatigue failure of whole bone under cyclic biaxial loading. The modeling pipeline was able to explain 85% of the variance in fatigue life when damage parameter A was calibrated and 91% when damage parameter B was calibrated. These strong predictive capabilities suggest that the modeling pipeline may provide valuable insights into the mechanisms underlying fatigue fracture in whole bone under multiaxial loading.

The CDM-based FE model used specimen-specific damage parameters for an accurate prediction of fatigue life, and our findings suggest the observed variability in damage parameters was driven by intrinsic bone properties rather than external mechanical factors. Indeed, the variability in damage parameters was not statistically different among loading groups, suggesting their independence from superposed torsion (Figure 2c, 2d). Bone material properties, particularly mineralization and porosity, strongly influence fatigue resistance. Higher mineralization and lower porosity are generally associated with increased stiffness and reduced damage accumulation, but this may come at the cost of reduced fracture toughness, limiting the bone's ability to resist crack propagation [42–44]. While some aspects of mineralization and porosity are captured through CT-based measurements of BMD, our CDM-based FE model cannot explicitly account for pre-existing microcracks or osteonal architecture, but the strong agreement between model predictions and experimental measurements suggests that CT captures many of the important aspects of fatigue resistance.

In this study, the damage parameters (A and B) were not simultaneously optimized (Section 2.1) or predicted (Section 2.2); instead, one parameter was held fixed while the other was calibrated for each model (Figure 2 and Figure 5). The benefit of this approach was that it enabled a more precise evaluation of the relationship between damage parameters and fatigue life. Interestingly, no statistically significant differences were found between the two calibration scenarios when comparing stress, strain, damage distributions (Figure 3) and compressive stiffness patterns (Figure 4). This suggests that the damage evolution and overall mechanical response of the bone were not highly sensitive to which parameter was fixed during calibration. The similar stress-strain and stiffness patterns indicate that both parameters contribute to fatigue behavior in a complementary manner; however, the superior predictive performance of ML models for parameter B suggests that it is more stable and sensitive for capturing fatigue behavior, leading to more accurate fatigue-life predictions. Unlike parameter A, which exhibited high variability (2.3–135), parameter B had a narrower range (2.5–3.6), making it easier for ML models to identify patterns. With a larger dataset, parameter A's variability may become more predictable, potentially improving ML accuracy for both parameters and further enhancing fatigue-life predictions.

Among the ML input features examined, "initial compressive stiffness" and "initial torsional stiffness" demonstrated the highest predictive power, likely due to their direct physical relevance to the mechanical response of the bone. Both features represent structural compliance under biaxial loading, effectively representing the overall mechanical integrity of the sample. These features likely encapsulate key aspects of bone quality and damage progression, making them strong indicators for predicting failure [45–47]. In contrast, BMC was likely less predictive than initial stiffness because it does not fully capture the progressive nature of continuum damage accumulation and structural failure. Although BMC reflects bone quantity, it does not capture information regarding geometry or mineral distribution that may influence fracture risk [48–50]. On the other hand, information related to geometry and mineral distribution is inherently captured with initial stiffness measurements.

Our CDM-based FE model requires information from ML for an accurate prediction of experimental measurements. This begs the question, can ML, without CDM-based FE modeling, be used to predict fatigue-life measurements? We tested this question post hoc, wherein ML models were trained using only initial compressive stiffness and initial torsional stiffness from FE analysis (Figure S2 in the **supplementary material**). The predictions showed poor accuracy ($R^2 \leq 0.09$). This poor predictive performance of ML predictions is likely due to the limited sample size of our study (n = 21), which restricts the ability of ML models to learn meaningful patterns between the input features and fatigue life. Given the inherent complexity and variability in bone fatigue behavior, a larger dataset would likely be required for ML to capture these relationships alone. In contrast, CDM-based FE models, with damage parameters informed by ML, demonstrated remarkable predictive capability (Figure 6), likely because they incorporate some of the fundamental mechanics governing fatigue failure. Unlike ML, which requires extensive data to generalize well, CDM-based FE models provide reliable insights even with a relatively small number of samples, making them particularly advantageous for studies where collecting a large dataset (e.g., > 200 samples) is considerably time-consuming [51–53].

This study has several limitations, not already mentioned, that should be born in mind when interpreting our results. We selected a 25% reduction in compressive stiffness as the failure criteria in line with previous work [33]; however, this may not be the ideal threshold for defining failure, especially under very high loads. For instance, some samples subjected to 50% torsion exhibited complete failure with less than a 3% reduction in stiffness. This might have influenced the optimization of the damage parameters for the group with 50% of the ultimate torsion in section 2.1.4, thereby driving the stiffness profile to the upper bound of experimental stiffness (Figure 4 at 50% torsion). This suggests that different failure criteria may be present in the low, medium, and high cycle regimes [54–57], which future studies can explore to potentially further improve the predictive capability of ML.

In conclusion, this study integrated Continuum Damage Mechanics (CDM)-based FE modeling and Machine Learning (ML) to predict fatigue life and stiffness degradation in whole bone under cyclic biaxial loading. Initial compressive and torsional stiffness were strong predictors of damage parameters. The CDM-based FE approach, informed by ML, effectively captured damage evolution and explained up to 91% of the experimental variance in fatigue-life measurements. The strong agreement between model predictions and experimental data highlights the potential of the CDM-based FE modeling approach to enhance our understanding of fatigue fracture mechanisms in whole bone.

**Acknowledgments**

This work was supported in part by the Natural Science and Engineering Research Council of Canada (NSERC; RGPIN 02404–2021 and RTI 00013–2016) and the Canadian Foundation for Innovation (CFI) John R. Evans Leaders Fund (JELF; Project #37134).

# Supplementary materials

**Continuum Damage Modeling of Biaxial Fatigue Failure in Whole Bone: A Hybrid Approach with Machine Learning Integration**

Reza Kakavand[1,2,3], Jonah M. Dimnik[1,2,3], Ifaz T. Haider[1,2,3], W. Brent Edwards[1,2,3]

[1] Human Performance Laboratory, Faculty of Kinesiology, University of Calgary, Calgary, Canada

[2] McCaig Institute for Bone and Joint Health, University of Calgary, Calgary, Canada

[3] Department of Biomedical Engineering, Schulich School of Engineering, University of Calgary

# 1 Isotropy and anisotropy

Isotropic FE models were described by $E_3$ only and a Poisson's ratio of $v=0.3$. Anisotropy was defined as described in section 2.1.2. Constant damage parameters of $A=10.40$ and $B=3.241$ from our previous study were considered for both isotropic and anisotropic CDM-based FE models [1].

Both experimental data and anisotropic CDM-based FE models showed a significant reduction in fatigue life with increasing superposed torsion ($p < 0.05$), while isotropic CDM-based FE models showed no significant differences among groups (Figure S1). Anisotropy explained 76% of the experimental variation in fatigue life compared to 39% for isotropy. Both models showed consistent fracture locations with the experiment. In pure compression, the fracture occurred at the proximal end, while with superposed torsion, the fracture occurred more distally. Both models overpredicted fatigue-life, highlighting the need for further calibration of the damage model.

CDM-based FE models of whole bone with anisotropic material properties demonstrated enhanced precision in replicating experimental biaxial fatigue-life measurements when compared to isotropic models. These findings emphasize the importance of incorporating anisotropy into our simulations of whole-bone biaxial fatigue failure.

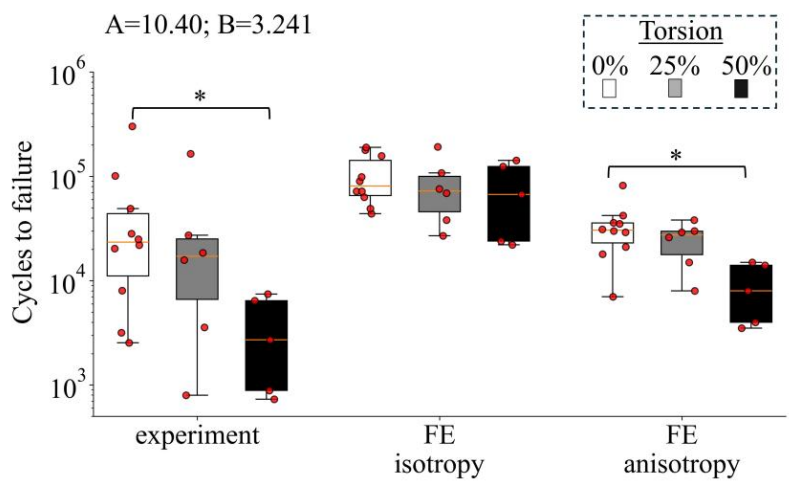

Figure S1. Fatigue-life of whole rabbit tibiae under 0%, 25%, and 50% of ultimate torsion (3.4 Nm) superposed on 50% of ultimate axial compression (2500 N). A and B from Dimnik et al [1].

## 2 Predicting fatigue-life directly from Machine Learning

Three ML models—Linear Regression, Random Forest (RF), and Support Vector Regression (SVR)—were implemented using their default parameters from the Scikit-Learn Python package. Given the limited dataset size (n = 21), leave-one-out cross-validation (LOOCV) was used as the model evaluation technique. In LOOCV, each sample is iteratively used as the test set while the remaining (n-1) samples serve as the training set. ML models were trained using initial compressive stiffness and initial torsional stiffness from FE analysis. ML performance was evaluated using the coefficient of determination ($R^2$), root mean squared error (RMSE) and mean absolute error (MAE). The predictions showed poor accuracy ($R^2 \leq 0.09$) as illustrated in Figure S2.

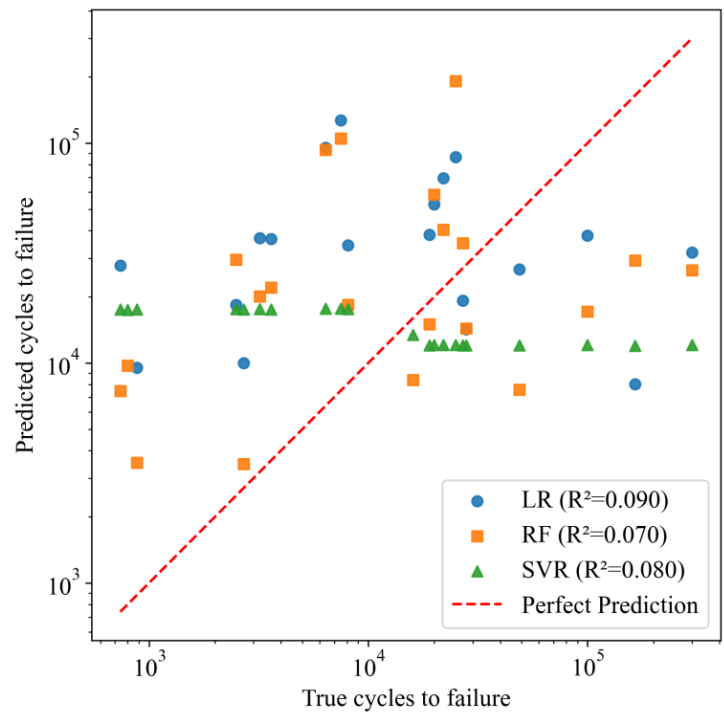

Figure S3. Machine Learning predicted verses experimentally measured cycles-to-failure.